# Short-period compact undulator (sCCU19) construction report.

Alexander Temnykh[1*] and Ivan Temnykh [2]
[1]CHESS, Cornell University, Ithaca, NY 14850, USA;
[2] Pine Hollow Auto Diagnostics, Pennsylvania Furnace, PA 16865, USA
* Corresponding author, E-mail: abt6@cornell.edu

## Abstract

Following the successful completion and testing of the sCCU-type undulator with 28mm period, we have designed, prototyped, constructed and bench tested another sCCU-type undulator with a much shorter 19mm-period magnetic structure. sCCU-type undulators are compact variable-gap devices with hydraulic-assist driver and innovative hybrid magnetic structure. The hydraulic system provides compensation of magnetic forces and ensures compactness of the design.   Owing to innovative magnetic structure, these undulators demonstrate magnetic peak field ~15% stronger compared to conventional hybrid-structure PM undulators.

The newly constructed undulator is 1.5 m long, 0.38m high, 0.21 m wide and weighs ~ 180 kg. Magnetic structure has 19 mm period and 6.5 mm minimum gap. The gap can be varied from 6.5mm to 80mm. At minimum (6.5mm) gap, the undulator demonstrated 0.87 T peak magnetic field.   Magnetic field measurements indicated satisfactory uniformity of the field and acceptable field integrals through the entire gap range. High reliability of undulator mechanics and the control system was confirmed by extensive, varied and prolonged testing.

## 1.  Introduction

Development of the sCCU19 undulator was motivated by the needs of the new HMF [1] CHESS beam line. Two undulators will operate at this beam line. First, sCCU28 will provide x-rays in energy range starting from 1.7keV. The second, sCCU19, has been optimized for maximum photon flux in energy range ~10keV (Table 1).

Refs. [2,3,4,5] describe in detail the design, principle of operation and bench test results of sCCU28. sCCU19 is identical to sCCU28, except a few minor differences due to the shorter magnetic structure period. These differences are discussed in the next section.

sCCU19 was tested on the CHESS ID4B beam line; results (Section 4) met the performance expectations.

Table 1. sCCU28 and sCCU19 general parameters

|  | sCCU 28 | sCCU 19 |
|---|---|---|
| Dimension: L x H x W [m] | 1.52 x 0.38 x 0.21 | |
| Weigh [kg] | 180 | |
| Minimum Gap [mm] | 6.5 | |
| Period [mm] | 28.4 | 19 |
| Peak Field [T] at minimum gap | 1.329 | 0.784 |
| K_max | 3.29 | 1.35 |
| E1 [keV] at Kmax | 1.7 | 9.6 |
| Max. Hydraulic pressure [Psi] | 931 | 358 |

## 2. Inventor model and drawing location.

sCCU19 3D model as well as production drawing can be found in Vault in folder "… 6097\6097-700 HMF Undulator\6097-704 SCCU-19 Assembly"; 3D model in file "6097-704 sCCU-19.iam"; complete set of drawings in "6097-704.idw".

### 2.1. Mechanical design notes.

Identical to sCCU28 [2], sCCU19 has a stationary frame, movable magnet arrays and hydraulic assist gap-controlling mechanism (Fig. 1).

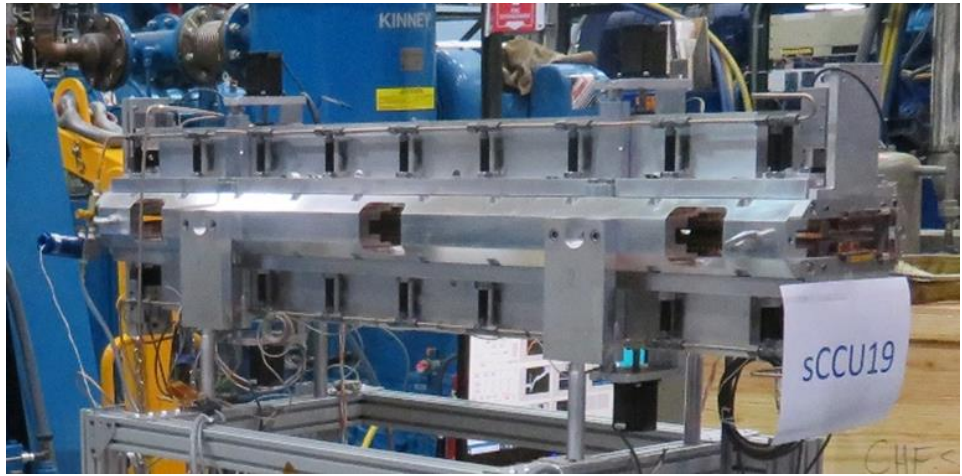

Figure 1. sCCU19 undulator in storage area

When the gap is small, the hydraulic system is enabled to compensate for the large attractive magnetic forces. The system consists of miniature hydraulic cylinders, hydraulic pump, pressure transducer etc. The Automatic Control System regulates hydraulic pressure depending on the gap.  Two mechanical drivers are installed to provide mechanical stability and precise positioning of the magnetic arrays. Operation principles of the hydraulic assist gap-controlling mechanism were described in [2,3,4].   All structural elements and components of the hydraulic system used for sCCU19 were identical to sCCU28. The only difference was the dependence of hydraulic pressure on gap, which was addressed by changing two parameters (m1 and m2) in the program loop (Fig. 2).

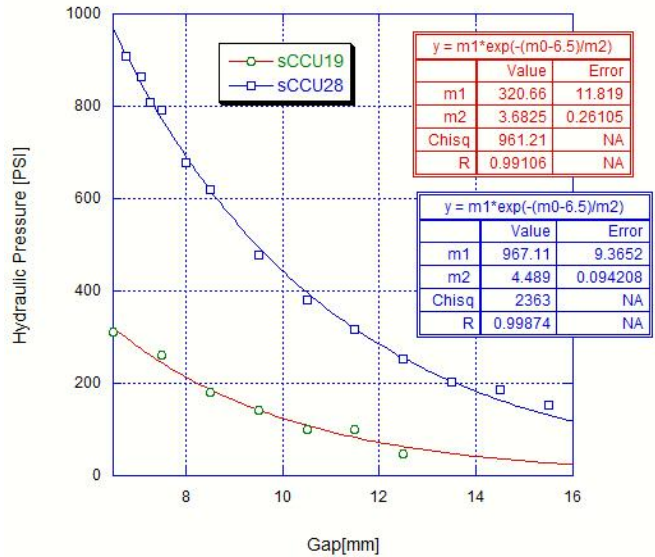

*Figure 2. Hydraulic pressure required for magnetic forces compensation for sCCU28 and sCCU19 undulators.*

Note that at minimum 6.5mm gap, the pressure required to compensate for the magnetic forces in sCCU28 is ~3.0 times (967psi/320psi) higher than it is for sCCU19. That is in good agreement with expectations, taking into account the quadratic scaling of magnetic forces with field amplitude: $(1.329T/0.784T)^2$ =~2.9.

### 2.2. Magnetic design and magnetic structure assembly notes.

sCCU19 magnet arrays are similar to the sCCU28 arrays: composed of single-pole assemblies with alternating magnetic field direction. Each pole assembly consists of a magnetic field concentrator made from Hiperco® 50 iron–cobalt–vanadium soft magnetic alloy attached to a copper holder (Fig. 3 Left) and 4 permanent magnet blocks with transfers dimensions [6] identical to sCCU28 blocks [2]. Due to the shorter period, the thickness of sCCU19 PM blocks was reduced from 4.75mm to 3.1mm. The small thickness prevented us from making grooves to fasten the PM blocks, so we glued them with epoxy after constraining the positions with stainless pins (red arrows, Fig. 3 Right).

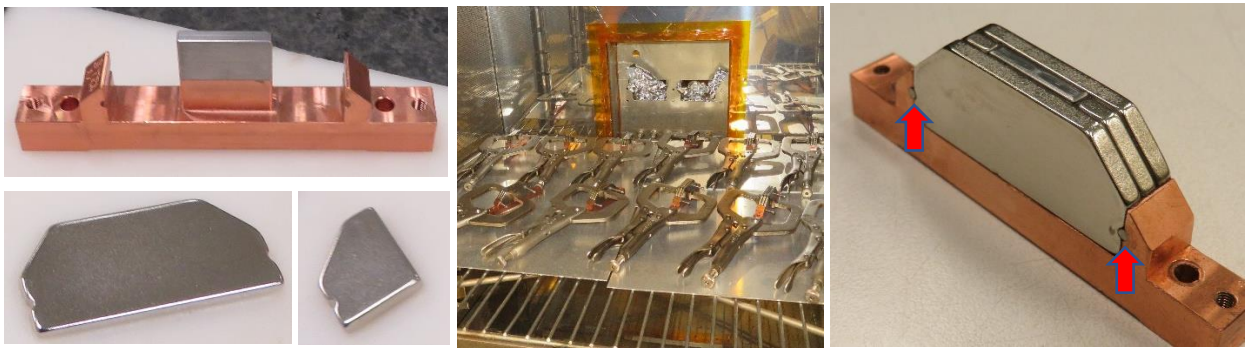

*Figure 3. sCCU19 magnetic structure assembly illustration. On the left – copper holders with attached field concentrators (top) and PM blocks. In the middle – assembled poles curing in oven (~2hrs at 55degC). On the right – assembled single pole.*

More information on magnetic structure assembly technique can be found in Ref.[6].

## 3. Assembled device characterization results

Magnetic field tuning of both arrays and then the assembled sCCU19 was performed in the same manner as for sCCU28 [2]. Final characterization results are presented below.

### 3.1. Peak field, "K" parameter and phase errors.

We measured the Peak Field and calculated RMS Errors and "Kund" as a function of gap (Fig. 4). At the minimum 6.5mm gap, sCCU19 produced 0.81 Tesla average peak field and 0.787 Tesla effective field corresponding to Kmax = 1.397. These numbers predict a first harmonic minimum energy of 9.11keV.

The measured RMS phase error varies from 3.7deg at minimum 6.5mm gap to 2.5deg at 11mm gap (Fig. 4 Left). This level of phase errors was considered acceptable because sCCU19 will operate at low order (1-st and 3-rd) harmonics.

A repeatability study of "Kund" parameter was performed (Fig. 4 Right). Here two "K" measurements (ramps 2 and 3) were taken as a function of undulator gap and their normalized difference (dK/K) was calculated. The "Kund" parameter showed virtually no changes between the two scans (dK/K ~0.001 with 0.0006 RMS), confirming very repeatable and stable operation of sCCU19.

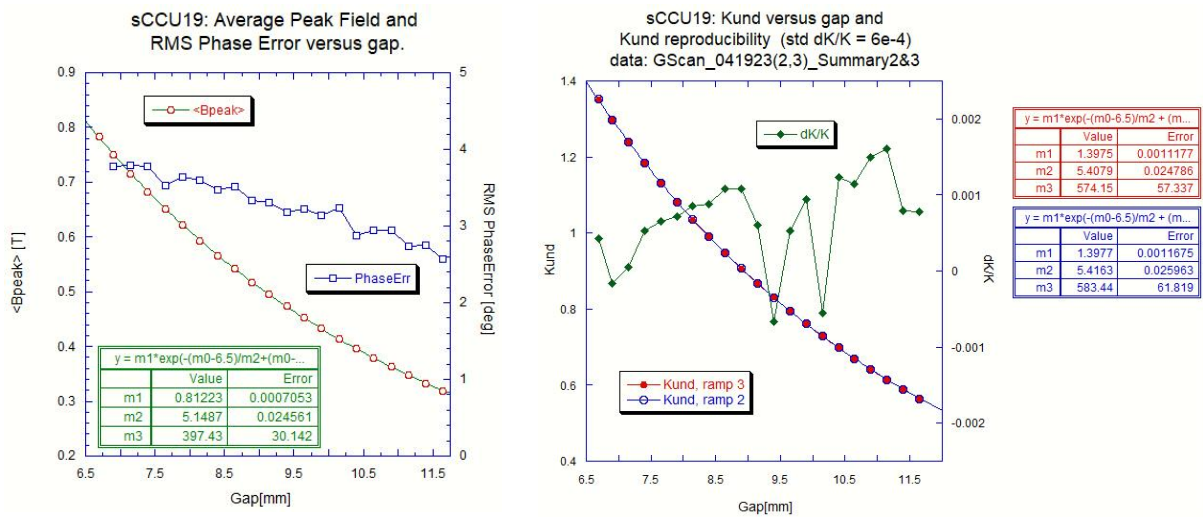

*Figure 4. On the left - sCCU19 peak field and RMS phase error measured as a function of gap. On the right – two measurements of "K" parameter as function of gap (left scale) and their normalized difference (right scale).*

### 3.2. Field Integrals variation and Magic Fingers (MF) correction.

Variation or dependence of ID field integrals on offset from beam axis may change beam dynamic in the storage ring. To avoid this, the variation of integrals should be minimized.

Dependence of vertical and horizontal field integrals (Ix, Iy) of sCCU19 on horizontal position was measured with long coils stretched through the undulator and corrected with the "magic fingers" technique as described in [2].

The data indicates that the MF correction procedure resulted in significant reduction of variation (Fig. 5). The "after correction" measurements for various gaps revealed a variation level typical for operating ID, ensuring that sCCU19 will not affect CESR beam dynamics when installed in the storage ring.

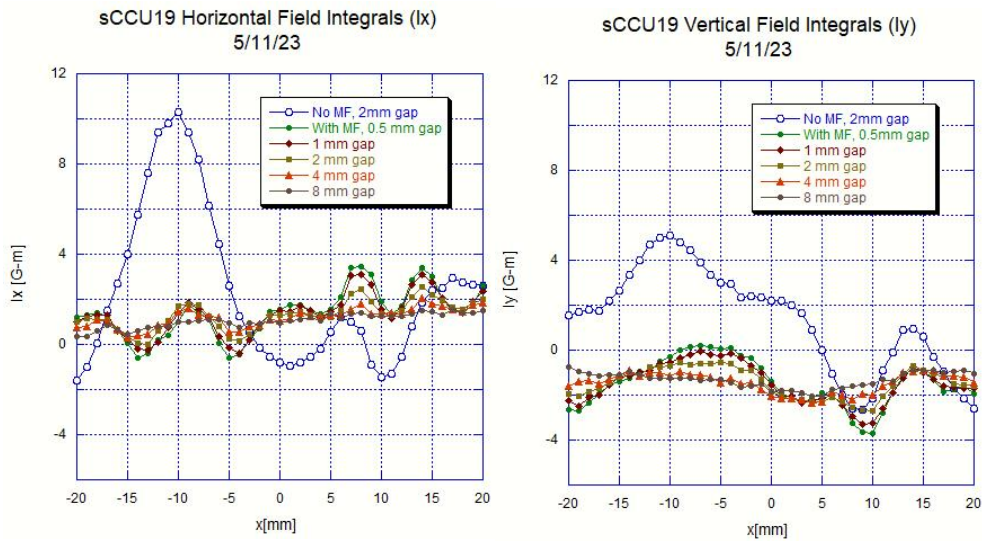

*Figure 5. Off axis horizontal (on the left)) and vertical (on the right) field integrals for various gaps. Open circles – indicate original integrals with no "magic fingers" correction, solid marks are for integrals with correction applied.*

### 3.3. End correction

Variation of the first and second field integrals while undulator gap is changing may result in undesired distortion of the closed orbit around the ring and displacement of the x-ray beams. To eliminate this variation, we installed two small electromagnets at the undulator's ends (end-correcting magnets). Each magnet can provide both vertical and horizontal field components to "zero" variation of the first and second field integrals.

Design, operation, calibration procedure, and successful performance tests of the end-correction magnets are described in Ref. [7].

## 4. Beam test results

In October 2023, the sCCU19 undulator was installed at Sector 4 in series with a standard CCU28 undulator (Fig. 6) and characterized using beamline ID4B instrumentation.

First, both undulators were tested at an energy range ~11keV. At this energy, sCCU19 radiates on the first harmonic (E1), while CCU28 must use the third (E3). In terms of flux, sCCU19 demonstrated ~2x higher flux than CCU28, in good agreement with the model (Fig. 7 Left).

In the energy range ~31keV, running on third harmonic (E3), sCCU19 demonstrated ~2.5x more flux than the standard CCU28, again in close agreement with the predicted values (Fig. 7 Right).

After this beam test confirmed undulator performance, sCCU19 was installed at Sector 5 to operate as a HMF beamline x-ray source.

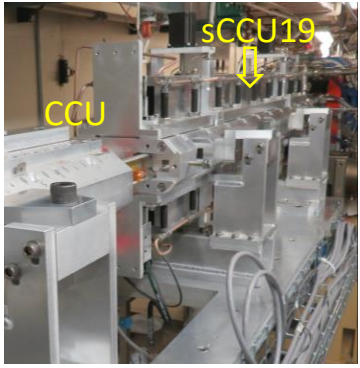
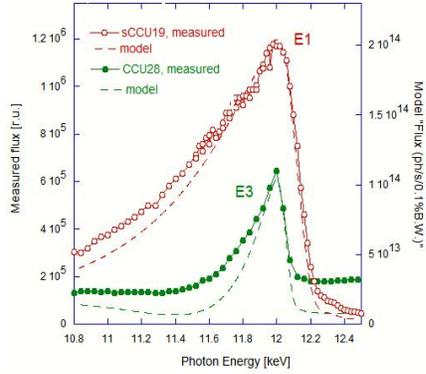
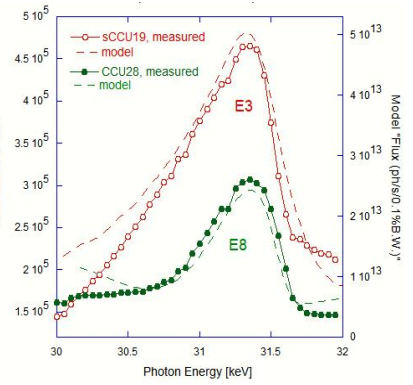

*Figure 6. Two undulators, nominal CCU and tested sCCU19, installed in CESR tunnel. October 2023.*

*Figure 7. Photon flux modeled and measured around 11keV (on the left) and 31keV (on the right). Red open circles indicate sCCU19 data, green solid is for CCU undulator.*

## 5. sCCU parameters scaling with period.

To facilitate the design process of future sCCU undulators, we developed an algorithm to optimize the magnetic structure period for the required photon energy range.

First, we plotted the measured magnetic field amplitude as a function of gap-to-period ratio for both sCCU19 (19mm period) and sCCU28 (28.4mm period) undulators. Remarkable data overlapping indicates that if we apply an appropriate fit, we can use it to predict the field amplitude for various gaps and for various periods (at least in the range from 19 to 28mm) (Fig. 8).

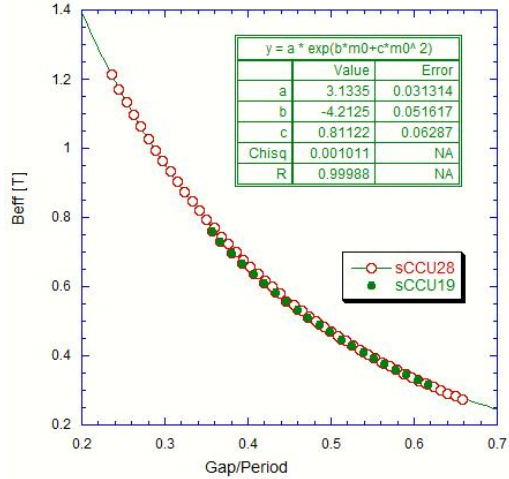

*Figure 8. Measured sCCU28 and sCCU19 effective magnetic field (Beff) as function of gap-to-period ratio.*

Following [8], the data in Fig. 8 was fitted with the expression:

$$B_{eff}[T] = a \times exp\left[b \times \left(\frac{g}{P}\right) + c \times \left(\frac{g}{P}\right)^2\right]$$

The fitting yielded: $a = 3.133 \pm 0.031$; $b = -4.21 \pm 0.05$; $c = 0.81 \pm 0.06$ with normalized residuals for $B_{eff}$ at ~0.006 level. Here: $g$ and $P$ are undulator gap and period in $mm$.

If we assume 6.5mm minimum undulator gap, the above expression will give dependence of field amplitude ($B_{eff}$) on period:

$$B_{eff}[T] = 3.13 \times exp\left[-\frac{27.43}{P} + \frac{38.06}{P^2}\right]$$

undulator parameter (K):

$$K_{und} = 0.292 \times P \times exp\left[-\frac{27.43}{P} + \frac{38.06}{P^2}\right]$$

and first harmonics energy ($E_1$) for 6GeV beam:

$$E_1[keV] = \frac{342}{P \times \left(1 + 0.043 \times P^2 \times exp\left[-\frac{54.86}{P} + \frac{76.12}{P^2}\right]\right)}$$

We can then plot [$B_{eff}$, $K_{und}$, $E_1$] as a function of sCCU period (6.5mm gap, 6GeV beam) (Fig. 9).

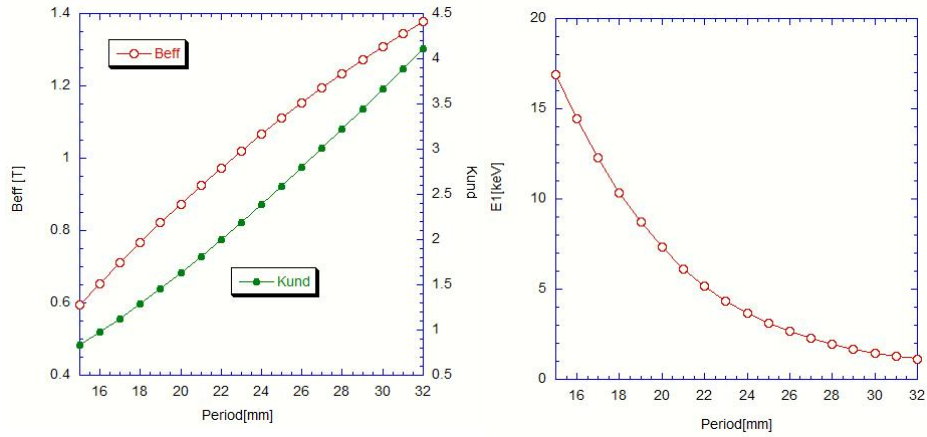

*Figure 9. On the left - effective field amplitude and Kund as function of sCCU period for 6.5mm gap. On the right – first harmonic energy as function of sCCU period for 6.5mm gap and 6GeV electron beam.*

## 6. Conclusion.

Two sCCU-type undulators with 19 and 28mm periods were constructed, characterized on the bench and installed in the storage ring for operation at the new HMF beamline. Prior to installation, sCCU19 was tested with electron beam in the storage ring. All results of both the characterization and beam test agreed very well with expectations.

In comparison to conventional variable-gap undulators with hybrid structure, sCCU-type undulators are more effective in terms of x-ray generation. They are also much more compact, lighter and cost-efficient.

## Acknowledgement

The authors would like to thank HMF beamline leading scientist Jacob Ruff for financing undulator construction. For design and fabrication of undulators parts we would like to thank Scott Hartman, Neil Alexander and Terry Neiss. Special thanks to accelerator technicians William Trask, Thomas Dugan and Ian Stilwell for help with permanent magnet blocks sorting and undulator assembly.